\documentclass[sigconf]{acmart}

\AtBeginDocument{%
  \providecommand\BibTeX{{%
    \normalfont B\kern-0.5em{\scshape i\kern-0.25em b}\kern-0.8em\TeX}}}

\setcopyright{acmcopyright}
\copyrightyear{2018}
\acmYear{2018}
\acmDOI{10.1145/1122445.1122456}

\acmConference[KDD-MLF '20]{KDD-MLF 2020: ACM SIGKDD Workshop on Machine Learning in Finance}{August 24, 2020}{San Diego, CA}

\begin{document}

\title{Machine Learning for Temporal Data in Finance:\protect\\ Challenges and Opportunities}

\author{ Jason Wittenbach}
\email{Jason.Wittenbach@capitalone.com }
\affiliation{
  \institution{Capital One - Center for Machine Learning}
  \streetaddress{1680 Capital One Drive}
  \city{McLean}
  \state{Virginia}
  \postcode{22102}
}

\author{ Brian d'Alessandro}
\email{brian.dalessandro@capitalone.com }
\affiliation{
  \institution{Capital One}
  \streetaddress{1680 Capital One Drive}
  \city{McLean}
  \state{Virginia}
  \postcode{22102}
}

\author{C. Bayan Bruss}
\email{Bayan.Bruss@capitalone.com}
\affiliation{%
  \institution{Capital One - Center for Machine Learning}
  \streetaddress{1680 Capital One Drive}
  \city{McLean}
  \state{Virginia}
  \postcode{22102}
}

\renewcommand{\shortauthors}{Wittenbach et. al.}

\begin{abstract}
Temporal data are ubiquitous in the financial services (FS) industry -- traditional data like economic indicators, operational data such as bank account transactions, and modern data sources like website clickstreams -- all of these occur as a time-indexed sequence. But machine learning efforts in FS often fail to account for the temporal richness of these data, even in cases where domain knowledge suggests that the precise temporal patterns between events should contain valuable information. At best, such data are often treated as uniform time series, where there is a sequence but no sense of exact timing. At worst, rough aggregate features are computed over a pre-selected window so that static sample-based approaches can be applied (e.g. number of open lines of credit in the previous year or maximum credit utilization over the previous month). Such approaches are at odds with the deep learning paradigm which advocates for building models that act directly on raw or lightly processed data and for leveraging modern optimization techniques to discover optimal feature transformations en route to solving the modeling task at hand. Furthermore, a full picture of the entity being modeled (customer, company, etc.) might only be attainable by examining multiple data streams that unfold across potentially vastly different time scales. In this paper, we examine the different types of temporal data found in common FS use cases, review the current machine learning approaches in this area, and finally assess challenges and opportunities for researchers working at the intersection of machine learning for temporal data and applications in FS.
\end{abstract}



\maketitle

\section{Introduction}
A core function of machine learning in financial services is modeling the behavior of customers or products and the markets in which they operate. For example, for consumer finance companies, the customers are credit card holders who operate in the debt market. Customer behavior and the market are both highly dynamic. At the same time, current and future states for both the customer and the market have strong temporal dependencies on their past states. As such, the data collected on them over the course of time should rightfully be viewed and modelled with a temporal component. 

\begin{figure}[h]
  \centering
  \includegraphics[width=\linewidth]{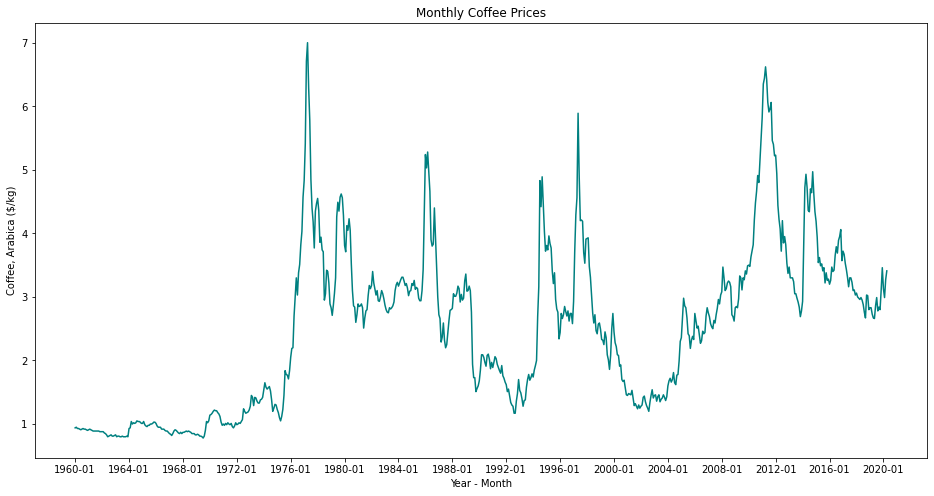}
  \caption{Finance has a long history of exploring and modeling time series of commodity prices as can be seen here with monthly prices of coffee dating back to the 1960s \cite{WB:2020}.}
  \Description{Coffee Commodity Prices}
\end{figure}

The history of modeling temporal data is intricately linked with the history of modern finance. Almost all investment is in some form based on an anticipation of a future state of the market as determined by current and historical knowledge. The most common of these is the futures contract where a producer of some commodity promises to deliver a good to the buyer at a specific time for a specific price. These types of contracts date back to the beginning of agriculture but found fertile soil in the 17th century Netherlands; the same birth place of the modern corporation. In these types of contracts, both the buyer and the seller need a model to anticipate the risk of the delivery of that good. These models have grown in sophistication along with the financial service industry and the technology that enables it \cite{petram2011}.

Until recently, much of the focus in finance around modeling temporal data has been limited to traditional data sources such as stock/commodity prices, deposits, and macroeconomic indicators. However, as FS companies digitize their operations, novel streams of data become available such as digital credit reports, clickstream data of customers online presence, detailed transaction information, and multi-channel payments. Each of these data streams can have a distinct time signature and scale and contain one or many different modalities of information. For instance clickstream data includes both the sequence of the session and time per page, but also the graph of the website, and the global point process of intents over sessions for that customer. 

Machine learning has seen significant advances in recent years in image classification and natural language processing. Within finance it has become a dominant tool for fraud detection and credit risk modeling. However, both in finance and in the broader machine learning research community, there remains a gap between methods developed for static data sets and those that can incorporate a temporal dimension. This paper presents a sample of recent advances in the relevant domains of machine learning on sequences and temporal data. The intent of this paper is to pose some challenges and opportunities for research into modeling of temporal data for financial applications.

\section{Recent Research}
At a high level, models of temporal data fall into the broader category of sequence models. Sequence models are those where the order of data points contains structural information. For example, language models are sequence models seeking to uncover the inherent structure to word choice, order, and dependency. Time series models are sequence models where there is explicit order and each data point is labeled with a time stamp. We could also call these event streams models to highlight that each observation will have a timestamp and a possible rich feature feature vector containing the details of the event. Because almost all industrial systems record the time at which the data was collected, nearly every dataset can be viewed as a temporal dataset. This can be further split into uniform and non-uniform time series, depending on whether the events/measurements are evenly spaced or come at irregular intervals. These data may come from discrete measurements of an underlying continuous process that take place at regular or irregular intervals (e.g. a stock price measured recorded each hour or a credit score requested only when a new application is submitted). Or the events may be singular, with no definition between observations (e.g. a credit card transaction). Much of the past research on time series forecasting has been in the domain of uniform time series. Non-uniform time series is a broad category containing a number of very common sequences in financial applications. One exception is when the focus is solely on the event timings, and not on any associated features or more complex tasks -- in this setting, the irregularly spaced events are called a point process, and there is an equally rich body on finance use cases in this area.

Each of these types of time series can be either univariate or multivariate. In the case of multivariate, the variables might be independent, but they may exhibit strong internal relationships at each time point. For example, video can be viewed as a multivariate time series of pixels. However, each frame contains strong internal structure and localized correlations which also interact along the time-axis. This can be viewed as a modality. There are time series with multiple modalities. Video with audio is an example where the sound and the images are distinct modalities with strong internal structures but relationships to each other and to the temporal dimension. Modalities can include images, text, graphs, and tabular feature spaces. 

\begin{figure}[h]
  \centering
  \includegraphics[width=\linewidth]{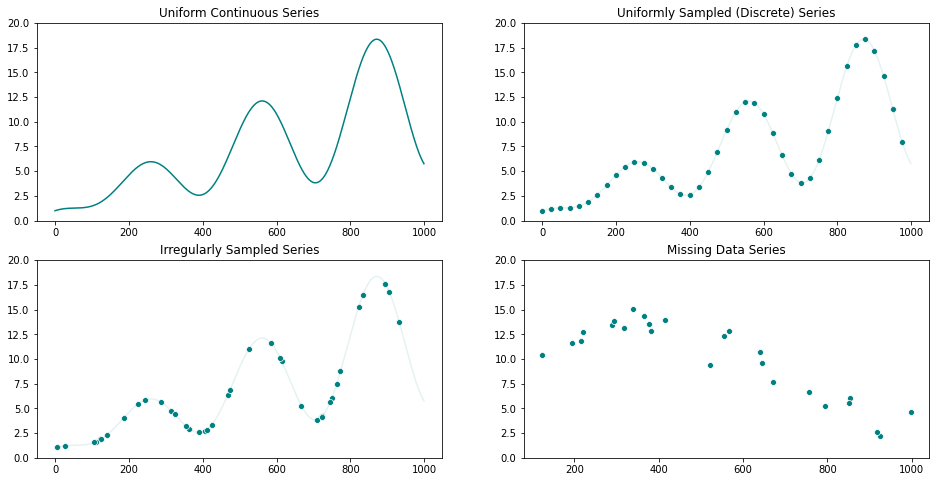}
  \caption{Examples of univariate time series with different profiles of sampling and missingness.}
  \Description{Different types of time series in FS}
\end{figure}

\subsection{Uniform Time Series}

Uniform time series can describe temporal data where the observations are evenly spaced in time. These time series can describe discrete processes or continuous processes sampled regularly. In finance, macroeconomic indicators such as unemployment rate and GDP are collected and reported on a regular cadence (weekly or monthly). This type of temporal data has received a tremendous amount of study over the last fifty years.  Much of this work has focused on univariate time series and building statistical techniques for decomposing varying components of the time series (e.g. level, seasonality, trend).  For a more in-depth overview of common time series methods please refer to \cite{hyndman2014forecasting}.

As shown in the recent Makridakis competition (M4), these statistical techniques still prove very robust for univariate uniform time series \cite{makridakis2018m4}.  At the time of the latest competition no pure machine learning approach had successfully surpassed the statistical benchmarks. However, for the first time the best performing model overall utilized a hybrid machine learning and statistical model. It is notable that this approach surpassed the baselines by nearly 10 percent. This model used a combination of a Holt-Winters statistical technique with a recurrent neural network \cite{smyl_2018}. The Holt-Winters multiplicative model decomposes a time series into three components: level, seasonality, and trend. The winning model observed that the assumption of linearity in the trend component in the Holt-Winters model might be effectively removed when you replace that component with a recurrent neural network. While the performance improvement is significant, this approach still requires heavy pre-processing of the data and carefully crafted neural architectures that depend on whether the series is monthly, quarterly or yearly. 

Oreshkin et. al. seek to build on this success by using a fully neural architectures for univariate time series. They establish three principles for a neural approach to time series:  1) a simple and generic architecture, 2) the architecture should not require heavy pre-processing and feature engineering of the time-series input and 3) it should be interpretable. The proposed approach uses stacks of residually connected layers that seek to learn basis functions to simultaneously predict the next value in the series while also looking backwards to predict the series up until that point \cite{Oreshkin2020}. This building block approach works well on its own but is not inherently interpretable. Using inductive bias it is easily modifiable to explicitly model common components such as seasonality and trend. This approach does well on the M4 benchmark, though it was not included in the competition.

\subsection{Non-Uniform Event Streams}
As previously mentioned, many use cases in modern FS leverage data that goes beyond uniformly sampled time series, instead dealing with data that are more properly described as irregular event streams -- sequences of feature vectors that are indexed by a continuous timestamp as opposed to an discrete integer. Examples include bank account transactions (withdrawals, deposits, transfers, etc) to click stream events recorded during a users interaction with a website. In cases where the precise timing of events contains task-relevant information, throwing away the timestamps and reverting to treating the data as a uniform time series will limit model performance. Fortunately, there is a rich and evolving literature around building deep learning models for even stream data.

\begin{figure}[h]
  \centering
  \includegraphics[width=\linewidth]{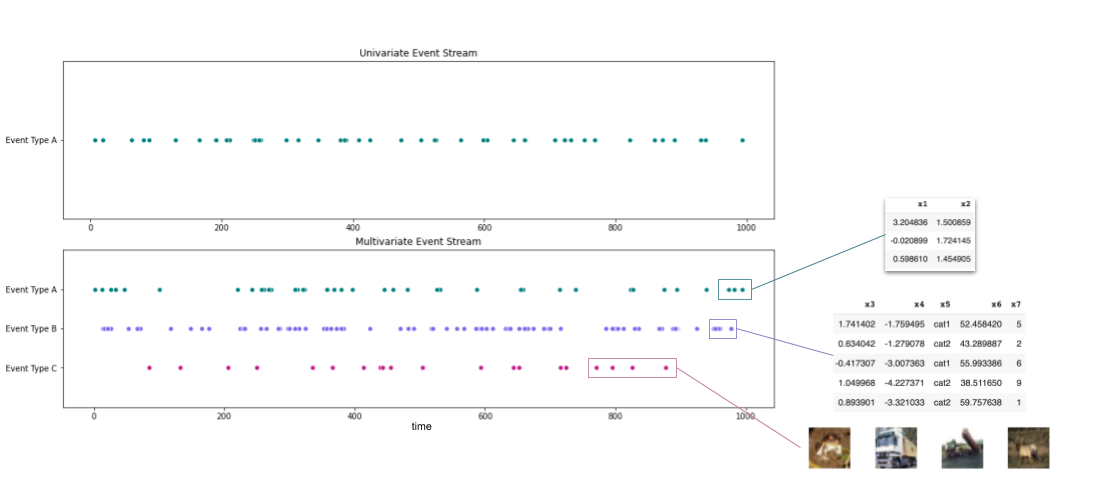}
  \caption{Examples of univariate and multivariate event streams. Individual events can be comprised of multiple features coming from possibly distinct modalities.}
  \Description{Different types of event streams}
\end{figure}

\subsubsection{Augmented Uniform Time Series Approaches}
One broad set of approaches to this challenge is to fall back on models designed to handle uniform time series -- namely recurrent neural network (RNN) architectures in the context of deep learning -- but preprocess the event stream data in such a way that the relevant information about event timing is encoded elsewhere.

For instance, one early approach was to discretize time into small enough bins so that one or no events occur within any single bin, and then treat the data as a uniform time series with missing feature vectors in bins with no events. Of course, this leads to a problem where many positions in the sequence will have ``missing’’ data, but any technique for handling missing data can now be used to bridge this gap \cite{little_statistical_2019}. This could be anything from simple imputation schemes that replace missing values with a default value, the mean, or the mode, to complex model-based approaches that try to impute the best value for the missing data from the non-missing data \cite{bengio_recurrent_1996, tresp_solution_1998, parveen_speech_2002}.

The downside of this approach is that, by covering over the fact that the data were missing, imputation gives the model no hint as to the the timing of the underlying events \cite{rubin_inference_1976}. A key observation is that, when the missing values arise from discretization of event stream data, the missingness is actually informative as to the event timing and thus may be relevant to the task. This led to a more refined set of approaches where the feature vector is augmented with a binary mask that indicates whether or not the rest of the feature vector is observed data or an imputation, which allows the model make use of the informative missingness \cite{lipton_directly_2016}. Taking this idea a step further, the feature vector can also be augmented with the time since the last observation to reinforce the temporal nature of the underlying data \cite{chen_neural_2018}.

Finally, information about event timing can also be embedded into the dynamics of the RNN itself. For example, this can be achieved by modifying RNN such that, between events, the hidden state decays exponentially toward some fixed point with a learnable decay rate\cite{chen_neural_2018}. The idea being that, as time passes, events further in the past fade from relevance and provide diminishing predictive power, and we move back to a state of ignorance. For some applications, this assumption may be appropriate, but in cases where long-range dependencies are important, it is likely to prove problematic. This approach can be further extended to richer event streams where the event has a feature vector along with each timestamp. One drawback of this approach is that the models it produces are somewhat limited to next-event prediction tasks and density estimation, and how to extend them to other sequence modeling tasks is not obvious.

\subsubsection{Point Process Approaches}
An alternative starting point to building deep learning models for event stream data is to eschew  discretization and instead confront the problem of modeling sparse temporal data head-on. This approach has a venerable history in the study of point processes -- stochastic models for datasets where each observation is simply a point in some space \cite{daley_introduction_2003}. Often, that space is the set of real numbers and the observations correspond to timestamps of a set of events; this is a temporal point process. But other prominent examples exist, such as spatial point processes used in geosciences for modeling the locations of earthquakes over a geographic region.

Just as parametric distributions exist for modeling various univariate data types (Gaussian distributions for real-valued data, Poisson distributions for count data, etc), there are classic temporal point process models, such as the Poisson process for independent events, the Hawkes process for self-excitatory event streams, and more. Often these models are specified by giving the conditional probability of the next event time given the preceding event history. One way to produce a more flexible point process model is to parameterize this conditional probability by a neural network rather than the classical approach of assuming a fixed functional form, which is the approach taken by the nueral Hawkes process \cite{du_recurrent_2016, mei_neural_2017}. These models apply to a broader class of data know as ``marked poisson processes'', which each timestamp has an associated feature vector. While these features vector could technically be defined over any space, in practice they are usually over a discrete set of categories, such that each category signifies a different event type. So while these models are appropriate for fusing multiple point processes (see below), they are more geared toward tasks that focus solely on predicting timing.

A recent arrival in this space that overcomes this obstacle is the neural ODE (ordinary differential equation) \cite{chen_neural_2018, rubanova_latent_2019}. This model arises from the observation that adding skip connections to the hidden layer in an RNN make the forward pass of the model identical to a discrete time approximation of a continuous ODE. Moving to a continuous time makes modeling event sequences relatively straightforward, since one can then model the feature vectors as observations that are conditional on the instantaneous value of the latent state at the event timestamp. Furthermore, extensions to this model allow events to treated instead as inputs to the model that can have an instantaneous effect on the hidden state (via an ODE-RNN hybrid). And the probability of an event occurring can be modeled as a time-dependent Poisson process, the rate of which is conditioned on the instantaneous latent state of the model. Thus, these models naturally many tasks and modeling assumptions about the relationships between event features and system state. Though new and relatively untested, neural ODE models are a promising and flexible approach to modeling event stream data.

\subsection{Multi-Modal Fusion}
Along with deriving the appropriate model for a time series, another area of interest is how to fuse multiple sources of temporal data when each source has not only its own time signature but also a distinct structure in the feature space. One area to look to in this regard is recent research in multi-modal fusion. This research primarily focuses on predictive tasks related to fusing video, audio, and natural language understanding/generation or some subset thereof. The challenge in this domain is that pixel space and word token space have strong internal dynamics but don't directly related to one another. In many cases multi-modal fusion seeks to learn joint representations of each modality for the purpose of a predictive task.

Early work in this area focused on learning joint representations of data coming from two or more modalities. In \cite{Karpathy_2015_CVPR} the authors seek to generate image descriptions by aligning word vectors to objects detected in the image. The aligned object-word pairs are used to generate natural language descriptions of the image as a whole. In \cite{zhu2015aligning} the goal is to determine alignments between movies (image \& subtitle) and the books on which they were based. Again this is an example of two modalities, text and images. However, now the image modality unfolds over time and the two text modalities follow different sequential patterns. In order to align the books and the movies they first generate sentence embeddings from the book while simultaneously generating scene embeddings of subtitled clips. Alignments between these two spaces are learned for those scenes that correspond to specific parts of the book. In \cite{10.5555/3016100.3016289} the alignments are learned between input instructions and sequences of spatial actions for a agent to execute. The goal of this is to be able to communicate with autonomous systems using spoken language.

As can be seen, much of the mulitmodal fusion work involves Recurrent Neural Networks (RNNs) to learn sequential dependencies in one or more of the streams of data. Perhaps unsurprisingly, the same efficacy that attention based models have shown for for a wide variety of language and graph tasks, has been observed in the multi-modal fusion domain. \cite{ShowAttend_Xu} use an attention mechanism to learn locations within an intermediate representation of an image to focus on for the generation of each word in a caption. This attention can be hard, meaning it selects a single location, or it can be a soft distribution over the entire image. \cite{hori2017attention} extends this approach beyond single images to include video and audio. \cite{NIPS2019_8297} takes a similar approach but here there are two attention mechanisms swapped across the images and the language domain. It is also a more generalized approach to learning joint representations of language and images. In all of these models, the data is often sequential but there is no conception of time beyond discrete ordered tokens. 

\section{Outstanding Challenges}

As described above, FS companies generally work with multiple streams of sequence data. Research in the topic of non-uniform time series gives direction on how to approach time series data where the non-uniformity is driven by natural or uneven sampling processes. A core challenge then is, and taking cues from the literature of multi-modal fusion, how does one combine two or more non-uniform time series that each has its own modality? In \cite{zhu2015aligning}, alignment of the modalities it the core modeling goal. In many FS applications, alignment would be a pre-processing step necessary to solving the core modeling task. While most FS sequences would be indexed by a real time dimension, naively combining them into a master sequence would produce extra sparse and irregular input vectors. 

If we take a componentized view to the fusion of multiple non-uniform data sequences, architecture complexity quickly adds up. With each sequence being unique in both its domain and time-irregularity, separate embedding and time treatment components would be needed for each. On top of that we would need components to fuse each modality into a single representation. With increasingly complex architectures, with more parameters and hyper-parameters to tune, training requirements could increase dramatically. While compute is often effectively managed in large FS companies (often by throwing more GPUs at the problem), time and data are certainly limiting factors. Even at the transactional scale of a top FS firm, data is fairly heterogeneous (in terms of types of customers, merchants and transaction details). A converged solution may be good on average, but could fail to best represent micro-segments of the population data (this is a problem in all machine learning, but one that is exacerbated by highly complex models).

Methods that result in increased complexity also pose specific risks to FS companies in terms of what is generally referred to as ``model governance.'' Models in FS companies are heavily regulated by several laws and statutes that set rules around credit underwriting, fraud detection, anti-money laundering, marketing, and data sharing. FS companies typically maintain compliance through multiple stages of internal governance. A lot of model governance focuses on model input features and model interpretability. Model complexity generally makes explanations more difficult, but emerging methods (e.g., LIME \cite{ribeiro__2016}, SHAP \cite{lundberg_unified_2017}) are gaining regulatory acceptance. As a fallback, models can be understood by the collection of input features (with maybe some measure of global feature importance). Features audits are particularly important in credit lending, as statutes like Fair Lending specifically blacklist features that may lead to discrimination in credit lending (and possible proxies for them). As FS companies adopt methods that operate on raw input streams, the notion of an explicit features fades away. Existing governance protocols are not designed for auditing implicit feature engineering, such as is done in deep learning. While deep learning may not pose a risk for including explicitly prohibited features (e.g., race, gender, sexual orientation), proxies to such features will be harder to detect when there are no explicit features to test. The solution to this challenge may not be in the deep learning models themselves, but coupling these models with auxiliary techniques for discrimination detection \cite{NIPS2016_6374} may have to be standard practice. 

\section{Conclusion}
The finance industry has a history of leading the way in the analysis of multivariate time series data, but recent trends in digitization have expanded both the amount and type of data available to companies in this sector. Classical time series are now joined by rich event streams, where observations with rich feature vectors take place at irregular intervals across multiple timescales and levels of organization. And streams are multiplying. A single bank account might contain multiple types of transaction events. A single customer might have multiple accounts as well as website interactions and credit bureau data on top of all of that. And it may very well be that the key piece of information for making a crucial business decision about that customer depends on understanding the interaction of the temporal patterns in those different streams. Across other industries, deep learning approaches have offered a way to build models that are tailored to the form and structure of the data and can excel at extracting meaningful patterns in order to solve a variety of tasks. Work has begun, but many challenges still exist for applying such models to complex and multimodal event streams. And once again, the finance industry is poised to lead the way.

\bibliographystyle{ACM-Reference-Format}
\bibliography{main}


\begin{thebibliography}{26}


\ifx \showCODEN    \undefined \def \showCODEN     #1{\unskip}     \fi
\ifx \showDOI      \undefined \def \showDOI       #1{#1}\fi
\ifx \showISBNx    \undefined \def \showISBNx     #1{\unskip}     \fi
\ifx \showISBNxiii \undefined \def \showISBNxiii  #1{\unskip}     \fi
\ifx \showISSN     \undefined \def \showISSN      #1{\unskip}     \fi
\ifx \showLCCN     \undefined \def \showLCCN      #1{\unskip}     \fi
\ifx \shownote     \undefined \def \shownote      #1{#1}          \fi
\ifx \showarticletitle \undefined \def \showarticletitle #1{#1}   \fi
\ifx \showURL      \undefined \def \showURL       {\relax}        \fi
\providecommand\bibfield[2]{#2}
\providecommand\bibinfo[2]{#2}
\providecommand\natexlab[1]{#1}
\providecommand\showeprint[2][]{arXiv:#2}

\bibitem[\protect\citeauthoryear{??}{WB:}{202}]%
        {WB:2020}
 \bibinfo{year}{202}\natexlab{}.
\newblock \bibinfo{title}{Coffee Prices}.
\newblock
\newblock
\urldef\tempurl%
\url{https://www.worldbank.org/en/research/commodity-markets#1}
\showURL{%
\tempurl}


\bibitem[\protect\citeauthoryear{Bengio and Gingras}{Bengio and
  Gingras}{1996}]%
        {bengio_recurrent_1996}
\bibfield{author}{\bibinfo{person}{Yoshua Bengio} {and}
  \bibinfo{person}{Francois Gingras}.} \bibinfo{year}{1996}\natexlab{}.
\newblock \showarticletitle{Recurrent neural networks for missing or
  asynchronous data}. In \bibinfo{booktitle}{\emph{Advances in neural
  information processing systems}}. \bibinfo{pages}{395--401}.
\newblock


\bibitem[\protect\citeauthoryear{Chen, Rubanova, Bettencourt, and
  Duvenaud}{Chen et~al\mbox{.}}{2018}]%
        {chen_neural_2018}
\bibfield{author}{\bibinfo{person}{Tian~Qi Chen}, \bibinfo{person}{Yulia
  Rubanova}, \bibinfo{person}{Jesse Bettencourt}, {and}
  \bibinfo{person}{David~K. Duvenaud}.} \bibinfo{year}{2018}\natexlab{}.
\newblock \showarticletitle{Neural ordinary differential equations}. In
  \bibinfo{booktitle}{\emph{Advances in neural information processing
  systems}}. \bibinfo{pages}{6571--6583}.
\newblock


\bibitem[\protect\citeauthoryear{Daley and Vere-Jones}{Daley and
  Vere-Jones}{2003}]%
        {daley_introduction_2003}
\bibfield{author}{\bibinfo{person}{Daryl~J. Daley} {and} \bibinfo{person}{David
  Vere-Jones}.} \bibinfo{year}{2003}\natexlab{}.
\newblock \bibinfo{booktitle}{\emph{An introduction to the theory of point
  processes. {Vol}. {I}. {Probability} and its {Applications}}}.
\newblock \bibinfo{publisher}{New York). Springer-Verlag, New York,}.
\newblock


\bibitem[\protect\citeauthoryear{Du, Dai, Trivedi, Upadhyay, Gomez-Rodriguez,
  and Song}{Du et~al\mbox{.}}{2016}]%
        {du_recurrent_2016}
\bibfield{author}{\bibinfo{person}{Nan Du}, \bibinfo{person}{Hanjun Dai},
  \bibinfo{person}{Rakshit Trivedi}, \bibinfo{person}{Utkarsh Upadhyay},
  \bibinfo{person}{Manuel Gomez-Rodriguez}, {and} \bibinfo{person}{Le Song}.}
  \bibinfo{year}{2016}\natexlab{}.
\newblock \showarticletitle{Recurrent marked temporal point processes:
  {Embedding} event history to vector}. In
  \bibinfo{booktitle}{\emph{Proceedings of the 22nd {ACM} {SIGKDD}
  {International} {Conference} on {Knowledge} {Discovery} and {Data}
  {Mining}}}. \bibinfo{pages}{1555--1564}.
\newblock


\bibitem[\protect\citeauthoryear{Hardt, Price, Price, and Srebro}{Hardt
  et~al\mbox{.}}{2016}]%
        {NIPS2016_6374}
\bibfield{author}{\bibinfo{person}{Moritz Hardt}, \bibinfo{person}{Eric Price},
  \bibinfo{person}{Eric Price}, {and} \bibinfo{person}{Nati Srebro}.}
  \bibinfo{year}{2016}\natexlab{}.
\newblock \showarticletitle{Equality of Opportunity in Supervised Learning}.
\newblock In \bibinfo{booktitle}{\emph{Advances in Neural Information
  Processing Systems 29}}, \bibfield{editor}{\bibinfo{person}{D.~D. Lee},
  \bibinfo{person}{M.~Sugiyama}, \bibinfo{person}{U.~V. Luxburg},
  \bibinfo{person}{I.~Guyon}, {and} \bibinfo{person}{R.~Garnett}} (Eds.).
  \bibinfo{publisher}{Curran Associates, Inc.}, \bibinfo{pages}{3315--3323}.
\newblock
\urldef\tempurl%
\url{http://papers.nips.cc/paper/6374-equality-of-opportunity-in-supervised-learning.pdf}
\showURL{%
\tempurl}


\bibitem[\protect\citeauthoryear{Hori, Hori, Lee, Zhang, Harsham, Hershey,
  Marks, and Sumi}{Hori et~al\mbox{.}}{2017}]%
        {hori2017attention}
\bibfield{author}{\bibinfo{person}{Chiori Hori}, \bibinfo{person}{Takaaki
  Hori}, \bibinfo{person}{Teng-Yok Lee}, \bibinfo{person}{Ziming Zhang},
  \bibinfo{person}{Bret Harsham}, \bibinfo{person}{John~R Hershey},
  \bibinfo{person}{Tim~K Marks}, {and} \bibinfo{person}{Kazuhiko Sumi}.}
  \bibinfo{year}{2017}\natexlab{}.
\newblock \showarticletitle{Attention-based multimodal fusion for video
  description}. In \bibinfo{booktitle}{\emph{Proceedings of the IEEE
  international conference on computer vision}}. \bibinfo{pages}{4193--4202}.
\newblock


\bibitem[\protect\citeauthoryear{Hyndman and Athanasopoulos}{Hyndman and
  Athanasopoulos}{2014}]%
        {hyndman2014forecasting}
\bibfield{author}{\bibinfo{person}{R.J. Hyndman} {and} \bibinfo{person}{G.
  Athanasopoulos}.} \bibinfo{year}{2014}\natexlab{}.
\newblock \bibinfo{booktitle}{\emph{Forecasting: principles and practice}}.
\newblock \bibinfo{publisher}{OTexts}.
\newblock
\showISBNx{9780987507105}
\urldef\tempurl%
\url{https://books.google.com/books?id=gDuRBAAAQBAJ}
\showURL{%
\tempurl}


\bibitem[\protect\citeauthoryear{Karpathy and Fei-Fei}{Karpathy and
  Fei-Fei}{2015}]%
        {Karpathy_2015_CVPR}
\bibfield{author}{\bibinfo{person}{Andrej Karpathy} {and} \bibinfo{person}{Li
  Fei-Fei}.} \bibinfo{year}{2015}\natexlab{}.
\newblock \showarticletitle{Deep Visual-Semantic Alignments for Generating
  Image Descriptions}. In \bibinfo{booktitle}{\emph{The IEEE Conference on
  Computer Vision and Pattern Recognition (CVPR)}}.
\newblock


\bibitem[\protect\citeauthoryear{Lipton, Kale, and Wetzel}{Lipton
  et~al\mbox{.}}{2016}]%
        {lipton_directly_2016}
\bibfield{author}{\bibinfo{person}{Zachary~C. Lipton}, \bibinfo{person}{David
  Kale}, {and} \bibinfo{person}{Randall Wetzel}.}
  \bibinfo{year}{2016}\natexlab{}.
\newblock \showarticletitle{Directly modeling missing data in sequences with
  rnns: {Improved} classification of clinical time series}. In
  \bibinfo{booktitle}{\emph{Machine {Learning} for {Healthcare} {Conference}}}.
  \bibinfo{pages}{253--270}.
\newblock


\bibitem[\protect\citeauthoryear{Little and Rubin}{Little and Rubin}{2019}]%
        {little_statistical_2019}
\bibfield{author}{\bibinfo{person}{Roderick~JA Little} {and}
  \bibinfo{person}{Donald~B. Rubin}.} \bibinfo{year}{2019}\natexlab{}.
\newblock \bibinfo{booktitle}{\emph{Statistical analysis with missing data}}.
  Vol.~\bibinfo{volume}{793}.
\newblock \bibinfo{publisher}{John Wiley \& Sons}.
\newblock


\bibitem[\protect\citeauthoryear{Lu, Batra, Parikh, and Lee}{Lu
  et~al\mbox{.}}{2019}]%
        {NIPS2019_8297}
\bibfield{author}{\bibinfo{person}{Jiasen Lu}, \bibinfo{person}{Dhruv Batra},
  \bibinfo{person}{Devi Parikh}, {and} \bibinfo{person}{Stefan Lee}.}
  \bibinfo{year}{2019}\natexlab{}.
\newblock \showarticletitle{ViLBERT: Pretraining Task-Agnostic Visiolinguistic
  Representations for Vision-and-Language Tasks}.
\newblock In \bibinfo{booktitle}{\emph{Advances in Neural Information
  Processing Systems 32}}, \bibfield{editor}{\bibinfo{person}{H.~Wallach},
  \bibinfo{person}{H.~Larochelle}, \bibinfo{person}{A.~Beygelzimer},
  \bibinfo{person}{F.~d\textquotesingle Alch\'{e}-Buc},
  \bibinfo{person}{E.~Fox}, {and} \bibinfo{person}{R.~Garnett}} (Eds.).
  \bibinfo{publisher}{Curran Associates, Inc.}, \bibinfo{pages}{13--23}.
\newblock
\urldef\tempurl%
\url{http://papers.nips.cc/paper/8297-vilbert-pretraining-task-agnostic-visiolinguistic-representations-for-vision-and-language-tasks.pdf}
\showURL{%
\tempurl}


\bibitem[\protect\citeauthoryear{Lundberg and Lee}{Lundberg and Lee}{2017}]%
        {lundberg_unified_2017}
\bibfield{author}{\bibinfo{person}{Scott~M. Lundberg} {and}
  \bibinfo{person}{Su-In Lee}.} \bibinfo{year}{2017}\natexlab{}.
\newblock \showarticletitle{A unified approach to interpreting model
  predictions}. In \bibinfo{booktitle}{\emph{Advances in neural information
  processing systems}}. \bibinfo{pages}{4765--4774}.
\newblock


\bibitem[\protect\citeauthoryear{Makridakis, Spiliotis, and
  Assimakopoulos}{Makridakis et~al\mbox{.}}{2018}]%
        {makridakis2018m4}
\bibfield{author}{\bibinfo{person}{Spyros Makridakis},
  \bibinfo{person}{Evangelos Spiliotis}, {and} \bibinfo{person}{Vassilios
  Assimakopoulos}.} \bibinfo{year}{2018}\natexlab{}.
\newblock \showarticletitle{The M4 Competition: Results, findings, conclusion
  and way forward}.
\newblock \bibinfo{journal}{\emph{International Journal of Forecasting}}
  \bibinfo{volume}{34}, \bibinfo{number}{4} (\bibinfo{year}{2018}),
  \bibinfo{pages}{802--808}.
\newblock


\bibitem[\protect\citeauthoryear{Mei, Bansal, and Walter}{Mei
  et~al\mbox{.}}{2016}]%
        {10.5555/3016100.3016289}
\bibfield{author}{\bibinfo{person}{Hongyuan Mei}, \bibinfo{person}{Mohit
  Bansal}, {and} \bibinfo{person}{Matthew~R. Walter}.}
  \bibinfo{year}{2016}\natexlab{}.
\newblock \showarticletitle{Listen, Attend, and Walk: Neural Mapping of
  Navigational Instructions to Action Sequences}. In
  \bibinfo{booktitle}{\emph{Proceedings of the Thirtieth AAAI Conference on
  Artificial Intelligence}} (Phoenix, Arizona)
  \emph{(\bibinfo{series}{AAAI’16})}. \bibinfo{publisher}{AAAI Press},
  \bibinfo{pages}{2772–2778}.
\newblock


\bibitem[\protect\citeauthoryear{Mei and Eisner}{Mei and Eisner}{2017}]%
        {mei_neural_2017}
\bibfield{author}{\bibinfo{person}{Hongyuan Mei} {and}
  \bibinfo{person}{Jason~M. Eisner}.} \bibinfo{year}{2017}\natexlab{}.
\newblock \showarticletitle{The neural hawkes process: {A} neurally
  self-modulating multivariate point process}. In
  \bibinfo{booktitle}{\emph{Advances in {Neural} {Information} {Processing}
  {Systems}}}. \bibinfo{pages}{6754--6764}.
\newblock


\bibitem[\protect\citeauthoryear{Oreshkin, Carpov, Chapados, and
  Bengio}{Oreshkin et~al\mbox{.}}{2020}]%
        {Oreshkin2020}
\bibfield{author}{\bibinfo{person}{Boris~N. Oreshkin}, \bibinfo{person}{Dmitri
  Carpov}, \bibinfo{person}{Nicolas Chapados}, {and} \bibinfo{person}{Yoshua
  Bengio}.} \bibinfo{year}{2020}\natexlab{}.
\newblock \showarticletitle{N-BEATS: Neural basis expansion analysis for
  interpretable time series forecasting}. In
  \bibinfo{booktitle}{\emph{International Conference on Learning
  Representations}}.
\newblock
\urldef\tempurl%
\url{https://openreview.net/forum?id=r1ecqn4YwB}
\showURL{%
\tempurl}


\bibitem[\protect\citeauthoryear{Parveen and Green}{Parveen and Green}{2002}]%
        {parveen_speech_2002}
\bibfield{author}{\bibinfo{person}{Shahla Parveen} {and} \bibinfo{person}{Phil
  Green}.} \bibinfo{year}{2002}\natexlab{}.
\newblock \showarticletitle{Speech recognition with missing data using
  recurrent neural nets}. In \bibinfo{booktitle}{\emph{Advances in {Neural}
  {Information} {Processing} {Systems}}}. \bibinfo{pages}{1189--1195}.
\newblock


\bibitem[\protect\citeauthoryear{Petram}{Petram}{2011}]%
        {petram2011}
\bibfield{author}{\bibinfo{person}{L. Petram}.}
  \bibinfo{year}{2011}\natexlab{}.
\newblock \showarticletitle{The world’s first stock exchange: how the
  Amsterdam market for Dutch East India Company shares became a modern
  securities market, 1602-1700}.
\newblock  (\bibinfo{date}{01} \bibinfo{year}{2011}).
\newblock


\bibitem[\protect\citeauthoryear{Ribeiro, Singh, and Guestrin}{Ribeiro
  et~al\mbox{.}}{2016}]%
        {ribeiro__2016}
\bibfield{author}{\bibinfo{person}{Marco~Tulio Ribeiro},
  \bibinfo{person}{Sameer Singh}, {and} \bibinfo{person}{Carlos Guestrin}.}
  \bibinfo{year}{2016}\natexlab{}.
\newblock \showarticletitle{" {Why} should i trust you?" {Explaining} the
  predictions of any classifier}. In \bibinfo{booktitle}{\emph{Proceedings of
  the 22nd {ACM} {SIGKDD} international conference on knowledge discovery and
  data mining}}. \bibinfo{pages}{1135--1144}.
\newblock


\bibitem[\protect\citeauthoryear{Rubanova, Chen, and Duvenaud}{Rubanova
  et~al\mbox{.}}{2019}]%
        {rubanova_latent_2019}
\bibfield{author}{\bibinfo{person}{Yulia Rubanova}, \bibinfo{person}{Tian~Qi
  Chen}, {and} \bibinfo{person}{David~K. Duvenaud}.}
  \bibinfo{year}{2019}\natexlab{}.
\newblock \showarticletitle{Latent {Ordinary} {Differential} {Equations} for
  {Irregularly}-{Sampled} {Time} {Series}}. In
  \bibinfo{booktitle}{\emph{Advances in {Neural} {Information} {Processing}
  {Systems}}}. \bibinfo{pages}{5321--5331}.
\newblock


\bibitem[\protect\citeauthoryear{Rubin}{Rubin}{1976}]%
        {rubin_inference_1976}
\bibfield{author}{\bibinfo{person}{Donald~B. Rubin}.}
  \bibinfo{year}{1976}\natexlab{}.
\newblock \showarticletitle{Inference and missing data}.
\newblock \bibinfo{journal}{\emph{Biometrika}} \bibinfo{volume}{63},
  \bibinfo{number}{3} (\bibinfo{year}{1976}), \bibinfo{pages}{581--592}.
\newblock
\newblock
\shownote{Publisher: Oxford University Press.}


\bibitem[\protect\citeauthoryear{Smyl}{Smyl}{2018}]%
        {smyl_2018}
\bibfield{author}{\bibinfo{person}{Slawek Smyl}.}
  \bibinfo{year}{2018}\natexlab{}.
\newblock \bibinfo{title}{M4 Forecasting Competition: Introducing a New Hybrid
  ES-RNN Model}.
\newblock
\newblock
\urldef\tempurl%
\url{https://eng.uber.com/m4-forecasting-competition/}
\showURL{%
\tempurl}


\bibitem[\protect\citeauthoryear{Tresp and Briegel}{Tresp and Briegel}{1998}]%
        {tresp_solution_1998}
\bibfield{author}{\bibinfo{person}{Volker Tresp} {and} \bibinfo{person}{Thomas
  Briegel}.} \bibinfo{year}{1998}\natexlab{}.
\newblock \showarticletitle{A solution for missing data in recurrent neural
  networks with an application to blood glucose prediction}. In
  \bibinfo{booktitle}{\emph{Advances in {Neural} {Information} {Processing}
  {Systems}}}. \bibinfo{pages}{971--977}.
\newblock


\bibitem[\protect\citeauthoryear{Xu, Ba, Kiros, Cho, Courville, Salakhutdinov,
  Zemel, and Bengio}{Xu et~al\mbox{.}}{2015}]%
        {ShowAttend_Xu}
\bibfield{author}{\bibinfo{person}{Kelvin Xu}, \bibinfo{person}{Jimmy~Lei Ba},
  \bibinfo{person}{Ryan Kiros}, \bibinfo{person}{Kyunghyun Cho},
  \bibinfo{person}{Aaron Courville}, \bibinfo{person}{Ruslan Salakhutdinov},
  \bibinfo{person}{Richard~S. Zemel}, {and} \bibinfo{person}{Yoshua Bengio}.}
  \bibinfo{year}{2015}\natexlab{}.
\newblock \showarticletitle{Show, Attend and Tell: Neural Image Caption
  Generation with Visual Attention}. In \bibinfo{booktitle}{\emph{Proceedings
  of the 32nd International Conference on International Conference on Machine
  Learning - Volume 37}} (Lille, France) \emph{(\bibinfo{series}{ICML’15})}.
  \bibinfo{publisher}{JMLR.org}, \bibinfo{pages}{2048–2057}.
\newblock


\bibitem[\protect\citeauthoryear{Zhu, Kiros, Zemel, Salakhutdinov, Urtasun,
  Torralba, and Fidler}{Zhu et~al\mbox{.}}{2015}]%
        {zhu2015aligning}
\bibfield{author}{\bibinfo{person}{Yukun Zhu}, \bibinfo{person}{Ryan Kiros},
  \bibinfo{person}{Rich Zemel}, \bibinfo{person}{Ruslan Salakhutdinov},
  \bibinfo{person}{Raquel Urtasun}, \bibinfo{person}{Antonio Torralba}, {and}
  \bibinfo{person}{Sanja Fidler}.} \bibinfo{year}{2015}\natexlab{}.
\newblock \showarticletitle{Aligning books and movies: Towards story-like
  visual explanations by watching movies and reading books}. In
  \bibinfo{booktitle}{\emph{Proceedings of the IEEE international conference on
  computer vision}}. \bibinfo{pages}{19--27}.
\newblock


\end{thebibliography}

\appendix

\end{document}